\documentclass[aps,pra,reprint,superscriptaddress,showpacs]{revtex4-1}
\usepackage{amsmath}
\usepackage{amssymb}
\usepackage{bbold}              
\usepackage{braket}
\usepackage{graphicx}
\usepackage{color}

\newcommand{\dd}{\text{d}}

\newcommand{\modulus}[1]{\left| #1 \right|}

\begin{document}

\title{An optimal single-electron charge qubit for solid-state double quantum dots}

\author{J. Mosakowski}
\email{jm668@cam.ac.uk}
\affiliation{Cavendish Laboratory, J. J. Thomson Avenue, CB3 0HE, Cambridge, United Kingdom}
\author{E. T. Owen}
\affiliation{Cavendish Laboratory, J. J. Thomson Avenue, CB3 0HE, Cambridge, United Kingdom}
\affiliation{London Centre for Nanotechnology, University College London, 17-19 Gordon Street, WC1H 0AH, London, United Kingdom}
\author{T. Ferrus}
\affiliation{Hitachi Cambridge Laboratory, J. J. Thomson Avenue, CB3 0HE, Cambridge, United Kingdom}
\author{D. A. Williams}
\affiliation{Hitachi Cambridge Laboratory, J. J. Thomson Avenue, CB3 0HE, Cambridge, United Kingdom}
\author{M. C. Dean}
\affiliation{Cavendish Laboratory, J. J. Thomson Avenue, CB3 0HE, Cambridge, United Kingdom}
\author{C. H. W. Barnes}
\affiliation{Cavendish Laboratory, J. J. Thomson Avenue, CB3 0HE, Cambridge, United Kingdom}
\date{\today}

\begin{abstract}
We report on an optimal single-electron charge qubit for a solid-state double quantum dot (DQD) system and analyse its dynamics under a time-dependent linear detuning, using GPU accelerated numerical solutions to the time-dependent Schr\"odinger equation.  The optimal qubit is found to have basis states defined as the symmetric and antisymmetric linear combinations of the lowest energy bonding and anti-bonding states of the DQD at zero bias.  In contrast to charge qubits defined by the two localised ground states of the uncoupled DQD, this choice of the basis causes the resulting dynamics to have a maximal overlap with an idealised two-state model.  Our optimal qubit basis states are not localised to a single quantum dot and, as such, initialising the qubit requires a particular sequence of gate pulses to take the system from an initial fiducial state of the DQD to the logical $0$ or $1$.  We determine this sequence using pulses that incorporate the expected experimental finite rise times.  We also show how to perform arbitrary single qubit operations on the Bloch sphere using spin-echo type pulsing, allowing us to obtain any qubit state with at most two single pulses.  Measurement of the optimal qubits is achieved by determining the probability of finding the electron in one of the dots.
\end{abstract}


\maketitle

\section{Introduction}

    Solid-state devices are attractive candidates for the implementation of practical quantum computation because of their high integrability into current industrial production lines as well as their cheap processing costs.  Owing to advances in growth and fabrication techniques, high scalability and long coherence times, significant advances have been made towards the realisation of practical, scalable qubits in silicon and III-V material-based quantum dot structures \cite{hayashi03,fujisawa06,petta05,petersson10,veldhorst14}.  Double quantum dots (DQDs), which consist of two adjacent tunnel-coupled quantum wells, offer a way of producing a qubit based on the charge of a single electron \cite{lloyd93,divincenzo00}.  In these systems, the qubit basis states are nominally considered to be the localised ground state wave functions of the uncoupled quantum dots.  Single qubit operations can be achieved by applying external potentials to surface gates \cite{petta05,petersson10} or nearby electron or hole reservoirs \cite{hayashi03,fujisawa06}.  These pulses alter the relative energies of the two quantum wells and the tunnel barrier between them, allowing the qubit to oscillate between the basis states.  Previous experimental work has realised manipulation of charge qubits in GaAs/AlGaAs \cite{hayashi03,dovzhenko11,kataoka09,petta05} and Si:P \cite{rossi10} devices.

    In general, experiments show that the dynamics of these charge qubits can be modelled approximately using a two-site localised state model (LSM) which obeys the Time Dependent Scr\"odinger Equation (TDSE)
    \begin{equation}
        \label{eq:TDSEeff}
        \hat{H}_\mathrm{eff}(t) \psi_\mathrm{eff}(t) = i \hbar \frac{\partial}{\partial t}\psi_\mathrm{eff}(t),
    \end{equation}
    with an effective Hamiltonian $\hat{H}_\mathrm{eff}$ defined by
    \begin{equation}
        \label{eq:Heff}
        \hat{H}_\mathrm{eff}(t) = \frac{1}{2} \,\varepsilon(t) \,\sigma_{z} - \frac{1}{2} \,\Delta \,\sigma_{x}  + \frac{1}{2}(E_B+E_{AB}).
    \end{equation}
    Here, $\varepsilon(t)$ is the `detuning parameter' which is pulsed in order to manipulate the charge qubit.  The constant term defines the energy zero, where $E_B$ and $E_{AB}$ are the energies of the bonding and antibonding states, i.e. the two lowest energy states, at $\varepsilon = 0$.  The parameter $\Delta$ is the `hybridisation energy' between the two localised states.  Ideally, the basis states of this model, labelled $\psi_L$ and $\psi_R$, should be time-independent and fully localised on each of the LSM sites.  However, for realistic potentials, the states fully localised on one side of a double dot potential contain contributions from high energy states which can cause complex intra dot charge oscillations to occur under charge qubit manipulation \cite{kataoka09}.  As we will see, the two lowest eigenstates of the system can be used as the basis states but since they are delocalised across the two dots, it is better to use orthogonal linear combinations that are localised so that readout of the qubit state can simply be a measurement of which dot the electron collapses to after manipulation.

    In this paper, we use a generic model effective potential for a solid-state DQD system to find an optimal qubit basis, which has greatest overlap with the stationary qubits across a range of detunings (section~\ref{sec:methods}).  In section~\ref{sec:qubitcontrol} we show how to initialise a single electron into one of the qubit basis states and how to perform a set of mutually orthogonal rotations on the Bloch sphere using finite rise-time pulses.  In section~\ref{sec:readout}, we provide a way of relating measurement outcomes to qubit superposition coefficients.  We finish with the discussion of the results (section~\ref{sec:discussion}) and the conclusions (section~\ref{sec:conclusion}).

\section{Optimal Single-Electron Charge Qubits}
    \label{sec:methods}
    \begin{figure}
    \includegraphics[width=\columnwidth]{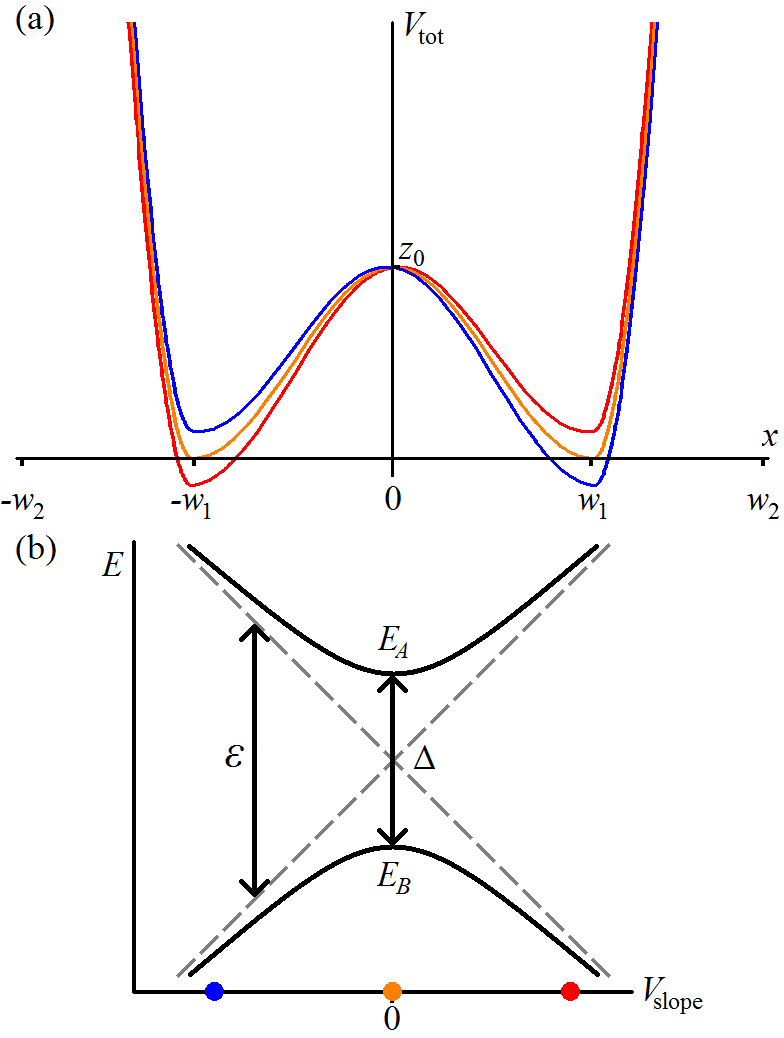}
    \caption{\label{fig:potmodel} (a) The DQD potential $V_{TOT}$ at zero (orange), lowest (blue) and highest (red) detuning values.  Their exact values are explained in Sec. \ref{subsec:optimalqubit}  (b) Energies $E$ of the bonding ($E_{B}$) and anti-bonding ($E_{A}$) eigenstates.  The coloured dots mark potential shapes from part (a).}
    \end{figure}

    We start by describing a four-parameter model for the effective potential in a DQD system that allows us to control both the depth of the dots and the barrier between them directly whilst also allowing us to apply detuning as a linear Stark shift.  Using this potential, we then define the optimal DQD qubit basis states being careful to make sure that the result is defined and accessible to experimental DQD systems.

\subsection{The double quantum dot potential}

    The effective potential in an experimental DQD system can be found using density functional theory \cite{owen15,stopa96} and will be a complex function of all three spatial coordinates $x,y,z$.  By careful design, the dynamics in two of the directions $y$ and $z$ can be confined to the lowest energy subbands so that only the potential in the $x$ direction, $V_{DQD}(x,t)$ needs be considered.  For example in a GaAs/AlGaAs heterostructure, the $z$ direction is the growth direction and modulation doping can be used to create a triangular quantum well in that direction with subband energies two orders of magnitude larger than either $\varepsilon$ or $\Delta$.  In the $y$ direction, parabolic confinement with energies an order of magnitude larger than $\varepsilon$ or $\Delta$ can be produced either by etching \cite{ferrus11}, fabricating a thin gate wrapping the conducting channel \cite{hisamoto90,voisin14} or using split-gates \cite{vanwees88}.  In order to create a DQD potential in the $x$ direction, gates \cite{fujisawa00,gardelis03,lim09,mason04} or etching \cite{wei13,ferrus11} can also be used.

    The aim is to create a potential $V_{DQD}(x,t)$ that has two minima separated by a tunnel barrier.  A convenient potential that has this property and is defined by four parameters $z_0$, $z_2$ and $w_1$, $w_2$ is given by
    \begin{equation}
        \label{eq:vdqd}
        V_{DQD}(x) = \begin{cases}\frac{1}{2} z_{0} \left[ 1 + \cos(\pi \frac{|x|}{w_{1}}) \right] & \text{if } |x| \leq w_{1} \\
        \frac{1}{2} z_{2} \left[ 1 - \cos(\pi \frac{|x|-w_{1}}{w_{2}-w_{1}}) \right] & \text{if } w_{1} < |x| \leq w_{2} \end{cases}
    \end{equation}
    This form for $V_{DQD}$ allows us to control both the depth of the dots and the barrier between them directly, using the parameters $z_{0}$ and $z_{2}$, respectively.  For a specific set of parameters, this static potential will define a value for $\Delta$ which is the energy difference between the bonding ground state $E_B$ and the antibonding first excited state $E_{AB}$.  Detuning is introduced by adding a linear stark shift of the form
    \begin{equation}
        \label{eq:vbias}
        V_\mathrm{bias} (x) = V_\mathrm{slope} \frac{x}{2 w_2}.
    \end{equation}
    By comparing the dependences of $E_B$ and $E_{AB}$ on $V_\mathrm{slope}$  with the expected dependences from LSM Hamiltonian we can define the detuning parameter for $V_{DQD}$ through a linear relation $\varepsilon = \lambda V_\mathrm{slope}$ with $\lambda$ being constant.  We find this linear relationship holds with an accuracy of one part in $10^6$ across the range of required values of $\varepsilon$ for single-qubit operations.   The total potential is $V_\mathrm{tot}(x) = V_\mathrm{DQD}(x) + V_\mathrm{bias}(x)$ and Fig.~\ref{fig:potmodel} (a) shows this potential at three different detunings.

    The DQD dynamics under time-dependent detuning will be given by the TDSE
    \begin{equation}
        \label{eq:TDSEreal}
        \hat{H}(x,t) \psi(x,t) = i \hbar \frac{\partial}{\partial t}\psi(x,t)
    \end{equation}
    with
    \begin{equation}
        \label{eq:realHamiltonian}
        \hat{H}(x,t) = - \frac{\hbar^2}{2 m^*} \frac{\partial^2}{\partial x^2} + V_{DQD} (x) + V_{\mathrm{bias}} (x,t).
    \end{equation}
    Time dependence is included in Eq. \ref{eq:TDSEreal} by varying the potential slope with time: $V_\mathrm{slope}(t)$.  An example plot of the energies of the two lowest instantaneous solutions (the bonding and antibonding states) as function of $V_\mathrm{slope}$ is shown in Fig.~\ref{fig:potmodel} (b).

    Analytic solutions to the TDSE in Eq. \ref{eq:realHamiltonian} can only be found in special cases.  In this paper we solve Eq.~\ref{eq:TDSEreal} numerically using a GPU-accelerated version of the staggered-leapfrog method~\cite{askar78, owen12} (see App.~\ref{subsec:ischeme}, App.~\ref{subsec:gpu}).

    Here and throughout the paper we use the parameter values: $w_{1} = 130 \text{nm}$, $w_{2} = 240 \text{nm}$, $z_{0} = 0.865 \text{meV}$, $z_{2} = 6.92 \text{meV}$ so that $\mathit{\Delta} = 12 \mu\text{eV}$, and the linear coefficient $\lambda = 0.42254$.  The values are chosen to be in agreement with the experiments~\cite{hayashi03, fujisawa04}, however, our conclusions are applicable to any DQD system and do not depend on the specific choice of parameter values.  We have also tested various non-symmetric potentials with the two dots having different sizes, but in all the cases the general conclusions were the same as for the symmetric potential of Eq. \ref{eq:vdqd}.

\subsection{The optimal qubit basis}
    \label{subsec:optimalqubit}

\begin{figure}
    \includegraphics[width=\columnwidth]{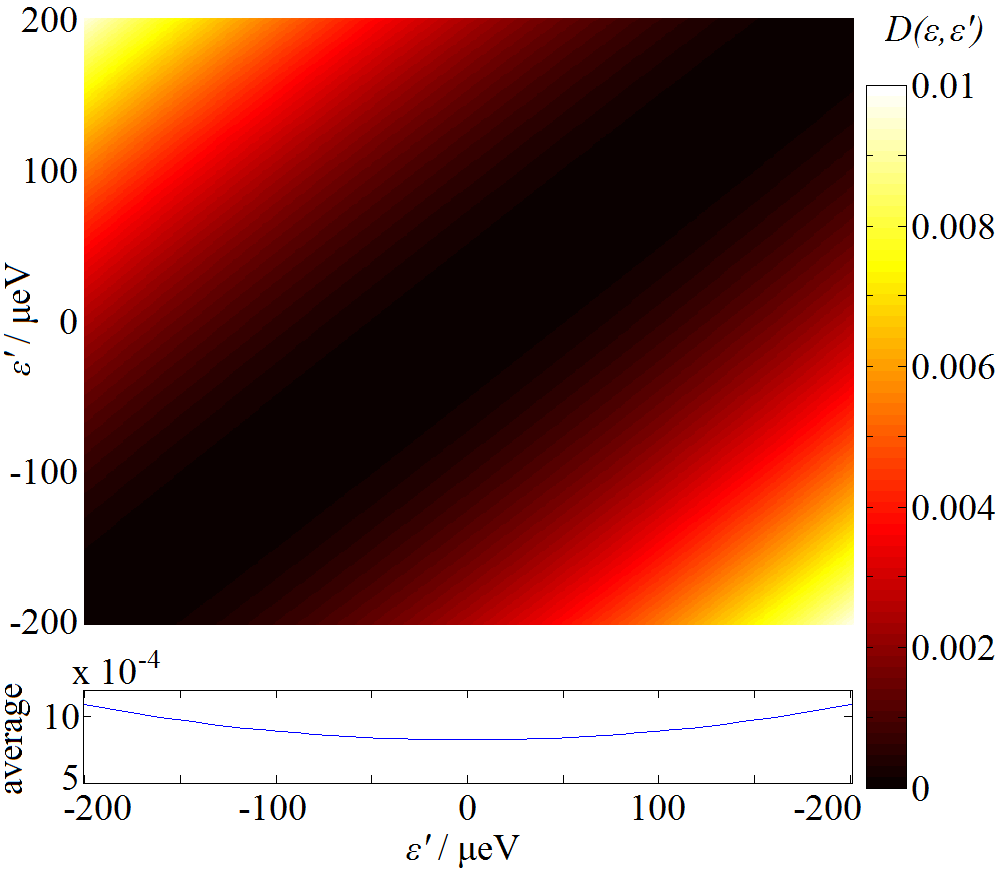}
    \caption{\label{fig:LocalisationOverlap} (Top) The average difference between the maximally localised states $D(\varepsilon, \varepsilon')$. (Bottom) Column averages of the top part of the plot.}
    \end{figure}

    We denote the optimal qubit basis for the potential $V_{DQD}(x)$ as the two states $\psi_0(x)$ and $\psi_1(x)$.   In order to make measurement of arbitrary qubit states $\psi(x)=\alpha \psi_0(x) + \beta \psi_1(x)$  as straightforward as possible these basis states would ideally be maximally localised in the left and right dots respectively.  This is because experimentally, it is possible to measure the probability that an electron will tunnel out of a quantum dot and this probability is in turn proportional to the probability that the electron is in that dot.  Also, in order to avoid complex transient oscillations, the qubit basis states should be linear combinations of the two lowest instantaneous eigenstates of the DQD system.  For the Hamiltonian in Eq.~\ref{eq:realHamiltonian}, the two lowest eigenstates are the instantaneous bonding and anti-bonding states $\psi_{\varepsilon}^B (x)$ and $\psi_{\varepsilon}^{AB} (x)$ for detuning $\varepsilon$.  Hence, the qubit basis states will be,
    \begin{eqnarray}
        \label{eq:maximallylocalisedR}
        R_\varepsilon (x) &=& \alpha_\varepsilon \psi_\varepsilon^B (x) + \beta_\varepsilon \psi_\varepsilon^{AB} (x), \\
        \label{eq:maximallylocalisedL}
        L_\varepsilon (x) &=& \beta_\varepsilon \psi_\varepsilon^B (x) - \alpha_\varepsilon \psi_\varepsilon^{AB} (x),
    \end{eqnarray}
    where $\alpha_\varepsilon$ and $\beta_\varepsilon$ are real variables,  $\alpha_\varepsilon^2 + \beta_\varepsilon^2 = 1$ and $\alpha_\varepsilon$ is chosen such that the integral
    \begin{equation}
        \int_0^\infty \modulus{R_\varepsilon (x)}^2 \dd x
    \end{equation}
    is maximised.  Readout would then simply be a matter of determining which dot the electron wave function collapsed into.  However, for a given pulse sequence these qubit basis states are time-dependent and cannot easily be compared to qubits with different detunings $\varepsilon$.  Therefore, it is preferable to define a stationary qubit basis, independent of $\varepsilon$.  Optimal basis states of this type should have the greatest overlap with $R_\varepsilon (x)$ and $L_\varepsilon (x)$ for all detunings.  Fig.~\ref{fig:LocalisationOverlap} shows a correlation function $D(\varepsilon, \varepsilon')$ that will be a minimum when this condition is met:
    \begin{equation}
        D(\varepsilon, \varepsilon') = 1 - \frac{ \int_{-\infty}^\infty \! \left( \modulus{L_\varepsilon (x) L_{\varepsilon'}(x)}^2 + \modulus{R_\varepsilon (x) R_{\varepsilon'}(x)}^2 \right) \! dx }{2}.
    \end{equation}
    The detuning range is chosen such that the fidelity
    \begin{equation}
        \int_{-\infty}^\infty \modulus{\phi_\varepsilon^*(x) \psi_\varepsilon^B(x)}^2 dx,
    \end{equation}
    is always $\geqslant 99\%$ for all values of $\varepsilon$, where, $\phi_\varepsilon(x) = a_\varepsilon R_0 (x) + b_\varepsilon L_0 (x)$, $(a_\varepsilon$,$b_\varepsilon$) is the corresponding bonding solution to $\hat{H}_\mathrm{{eff}}$ in Eq. \ref{eq:Heff}, and a similar measure is performed for the anti-bonding state.  As expected, $D(\varepsilon, \varepsilon')$ is smallest between states where the difference in detunings is small.  However, as we want to apply large biases across the double dot system, the best qubit states are the ones where $D(\varepsilon, \varepsilon')$ is minimised over the entire domain $\varepsilon - \varepsilon'$.

\begin{figure}
    \includegraphics[width=\columnwidth]{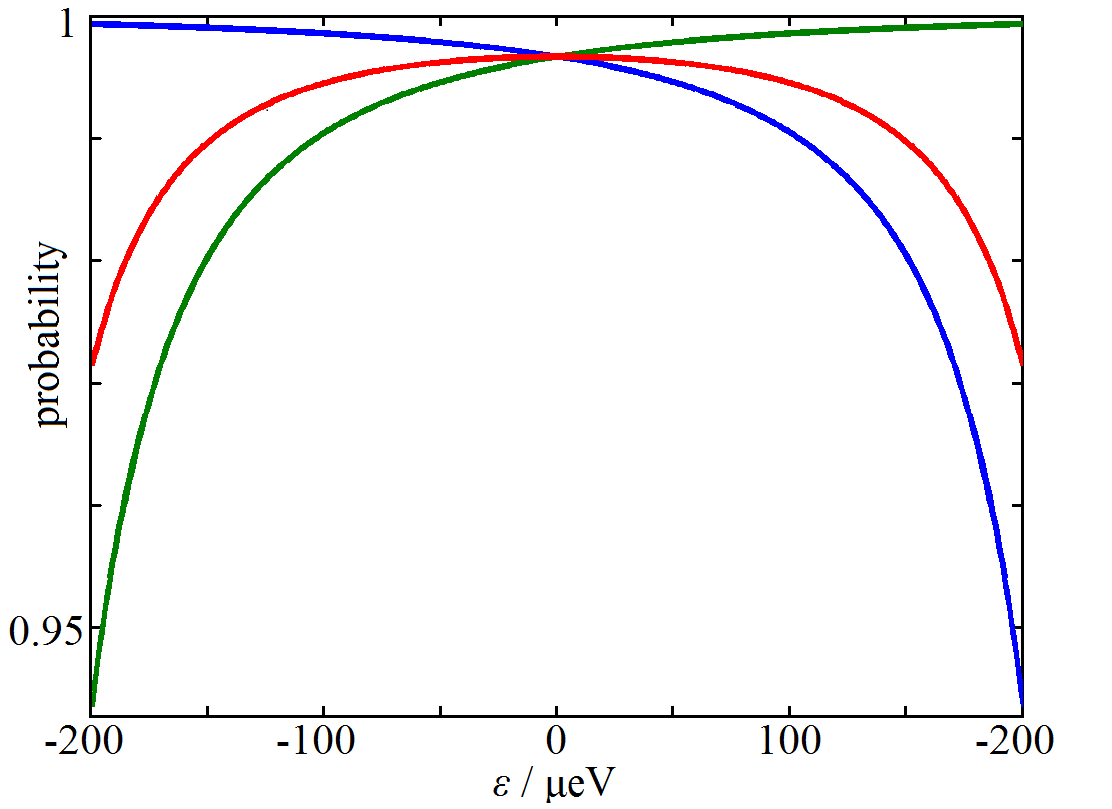}
    \caption{\label{fig:pRpLpSum} Probability of finding the particle in the left (blue) or right (green) dot for the maximally-localised left and right states, respectively, and the average of the two probabilities (red).}
    \end{figure}

    Using this criterion, we find that the best choice of states from which to build the qubits of the system are the bonding and anti-bonding states at zero detuning.  By parity inversion symmetry when $V_{\mathrm{bias}} (x, t) = 0$, if $x \to -x$ then $R_0(x) \to L_0(x)$ with $\psi_0^B (x) \to \psi_0^B (x)$ and $\psi_0^{AB} (x) \to - \psi_0^{AB} (x)$ so the coefficients $\alpha_0 = \beta_0 = 1 / \sqrt{2}$.  Therefore, our optimal charge qubits are defined as
    \begin{eqnarray}
        \psi_0(x) &= &R_0(x) =\frac{\psi_0^B (x) + \psi_0^{AB} (x)}{\sqrt{2}},\\
        \psi_1(x) &=& L_0(x)=\frac{\psi_0^B (x) - \psi_0^{AB} (x)}{\sqrt{2}}
    \end{eqnarray}
    These qubits are optimal in the sense that they have the greatest overlap with the maximally localised states in Eqs.~\ref{eq:maximallylocalisedR} and~\ref{eq:maximallylocalisedL} across the specified range of detunings.  These states are not fully localised as the probability density of $\psi_0(x)$ in the left dot is non-zero, and similarly for $\psi_1(x)$ in the right dot (Fig. \ref{fig:pRpLpSum}).  However, these are the states which maximise the sum of probabilities (red colour plot in Fig. \ref{fig:pRpLpSum}) and further localisation of the states would include higher-energy states which would not obey the ideal LSM Hamiltonian in Eq.~\ref{eq:Heff} which we want to model.

    The qubit we have established here, as a linear combination of bonding and anti-bonding states at zero detuning, is well defined in any DQD system including potentials that are not spatially symmetric.  Since, in the definition, there is no reference to the underlying effective potential of the DQD, this qubit is also well defined for an experimental system.

\section{Single Qubit Control}
    \label{sec:qubitcontrol}

    The qubit states we have chosen $\psi_0(x)$ and $\psi_1(x)$ form a static two-state system.  Their ideal dynamics is described by the TDSE with Hamiltonian $\hat{H}_\mathrm{eff}$ in Eq.~\ref{eq:Heff} so that, in terms of the polar and azimuthal angles on the Bloch sphere
    \begin{equation}
        \label{eq:qubitAngles}
        \phi_\mathrm{eff}(x,t) = \cos \! \left( \frac{\theta(t)}{2} \right) \psi_0(x) + e^{-i \phi(t)} \sin \! \left (\frac{\theta(t)}{2} \right)\psi_1(x).
    \end{equation}
    The dynamics of the state under $\hat{H}$ will be different but, we find that owing to the choice of the detuning range we have made (Sec. \ref{subsec:optimalqubit}), the difference between the solutions to Eq. \ref{eq:TDSEeff} and Eq. \ref{eq:TDSEreal} is always less than one percent.

    \begin{figure}
    \includegraphics[width=\columnwidth]{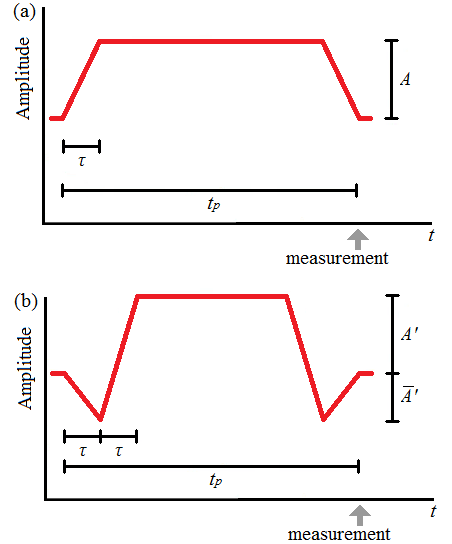}
    \caption{\label{fig:pulsing} (a) Standard and (b) spin-echo pulse types. To maximise the qubit probability value (before the electron decoheres), the measurement point was chosen to be just after the pulse ends \cite{comment1}.}
    \end{figure}

    A general qubit operation can be defined as a right-hand rotation on the Bloch sphere by an angle $\vartheta$ around a direction $\vec{n}$.  For the Hamiltonian in Eq.~\ref{eq:realHamiltonian}, this rotation is given by the solution to the TDSE:
    \begin{equation}
        \label{eq:arbitraryRotation}
        R_{\vec{n}} (\vartheta(t)) = \mathcal{T} \exp \left[ \frac{1}{i\hbar} \int_{0}^{t} \hat{H}(t') \dd t' \right]
    \end{equation}
    where $\mathcal{T}$ is the time-ordering operator.  We can perform the rotations by varying the detuning $\varepsilon$.  The simplest way to do this is by pulsing between two different values of the detuning $\varepsilon_0$ and $\varepsilon_1$ with a pulse duration $t_p$.  It is desirable for the pulse to instantaneously switch between $\varepsilon_0$ and $\varepsilon_1$ as this simplifies the dynamics and avoids spurious qubit rotations~\cite{koppens06}.  For an instantaneously pulsed Hamiltonian where the detuning $\varepsilon (t)$ can be described as a set of step-functions, $R_{\vec{n}} (\vartheta)$ can be expressed analytically as a rotation of the qubit state around the axis on the Bloch sphere which passes through the eigenstates of $H (t')$ at a rate proportional to the difference in energy of these two eigenstates.  Such a pulse requires a linear potential along the axis of the double dot potential, as in Eq.~\ref{eq:realHamiltonian}, which is achieved by applying voltages to a set of metallic surface gates.  However, the response of the electronics to an instantaneous pulse has a finite bandwidth which reduces the response time of the potential $V_{\mathrm{slope}} (t)$ and introduces a finite rise time $\tau$ (see Fig.~\ref{fig:pulsing} (a)).  In this case, the step-function decomposition is not possible and, in general, Eq.~\ref{eq:arbitraryRotation} must be solved numerically.  If the detuning can be described in terms of linear ramp functions, then Eq.~\ref{eq:arbitraryRotation} can be written analytically as a Landau-Zener-Stuckelberg transition~\cite{landau32, zener32, stueckelberg32} but the resulting expression is a function of parabolic cylinder functions which does not simplify understanding of the rotation $R_{\vec{n}} (\vartheta)$.  Instead, we solved Eq.~\ref{eq:arbitraryRotation} numerically using a GPU-accelerated version of the staggered-leapfrog method~\cite{askar78, owen12} (see App.~\ref{subsec:ischeme}, App.~\ref{subsec:gpu}).  In our simulations, we use a value of $\tau = 90$ps which is consistent with the experiments of Fujisawa \emph{et al.}~\cite{hayashi03, fujisawa04}.

    A generic feature of qubit rotations with finite rise times is that the path of an individual qubit state on the Bloch sphere during the time-evolution in Eq.~\ref{eq:arbitraryRotation} is not the same as the overall rotation given by $R_{\vec{n}} (\vartheta)$ ~\cite{foletti09}.  In fact, as with all solid rotations on a sphere, in order to describe $R_{\vec{n}} (\vartheta)$ we need to evolve three non-linear points on the sphere in order to reconstruct $R_{\vec{n}} (\vartheta)$:  For the Bloch sphere, this means three linearly independent states, which we chose to be $\psi_0(x), (\psi_0(x) + \psi_1(x)) / \sqrt{2}$ and $(\psi_0(x) + i \psi_1(x)) / \sqrt{2}$.

\subsection{State preparation}
    \label{sec:state_preparation}

    It is important to be able to initialise a qubit in a well defined state, preferably $\psi_0(x)$ or $\psi_1(x)$.  For a generic experiment involving a charge qubit, we would expect the initial state of the electron to be the ground state of Eq.~\ref{eq:realHamiltonian}.  This is not one of the qubit states so, in order to initialise the electron in the qubit state, we need to perform a rotation.  The exact parameters required for this procedure will vary for different detunings $\varepsilon$, for example, at certain detunings, the ground state wave function will be closer to one of the qubit basis states, but the general procedure reported here is the same.

    To rotate the wave function from the initial eigenstate to the qubit $\psi_0(x)$ state, we applied a trapezoidal pulse, as shown in Fig. \ref{fig:pulsing} (a). We performed a sweep of pulse lengths $t_p$ and detuning amplitudes $A = \varepsilon$ and calculated the distance
    \begin{equation}
        \label{eq:Sdistance}
        S(\psi_0, \psi) = 1 - \modulus{\int_{-\infty}^{\infty} \psi_0^*(x) \psi(x) dx}^2
    \end{equation}
    between the resultant state $\psi(x)$ and the qubit state $\psi_0(x)$.  For an initial static detuning $V_\mathrm{slope} (0) = 0.06508 \text{meV}$, corresponding to $\varepsilon = 27.5 \mu\text{eV}$ we found that this distance was minimized when $t_p = 537$ps and $A = 11.5 \mu$eV with an accuracy of $S < 10^{-5}$.

    In experiment, these parameters can be found by applying the same pulse to two distinct systems: one initialised in the ground state, and the other in the first excited state, then performing a sweep over pulse amplitudes and lengths and finding those which maximise the summed probability of finding the former state in the right and the latter one in the left dot, respectively, in agreement with Fig. \ref{fig:pRpLpSum}.

\subsection{$\sigma_x$ rotations}
    \label{subsec:sigmax}

    \begin{figure}
    \includegraphics[width=\columnwidth]{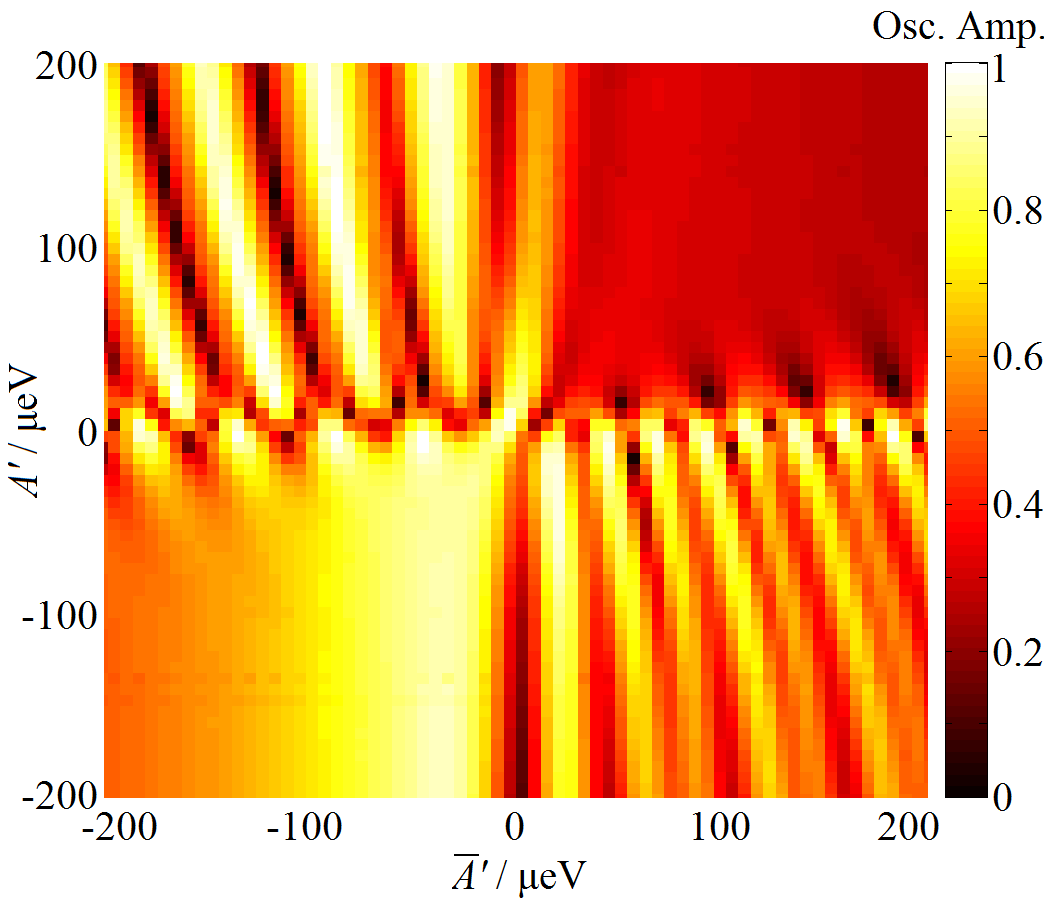}
    \caption{\label{fig:nmr} Qubit oscillation amplitude as a function of detunings $\bar{A}'$ and $A'$ for a spin-echo pulse type.}
    \end{figure}
    \begin{figure}
    \includegraphics[width=\columnwidth]{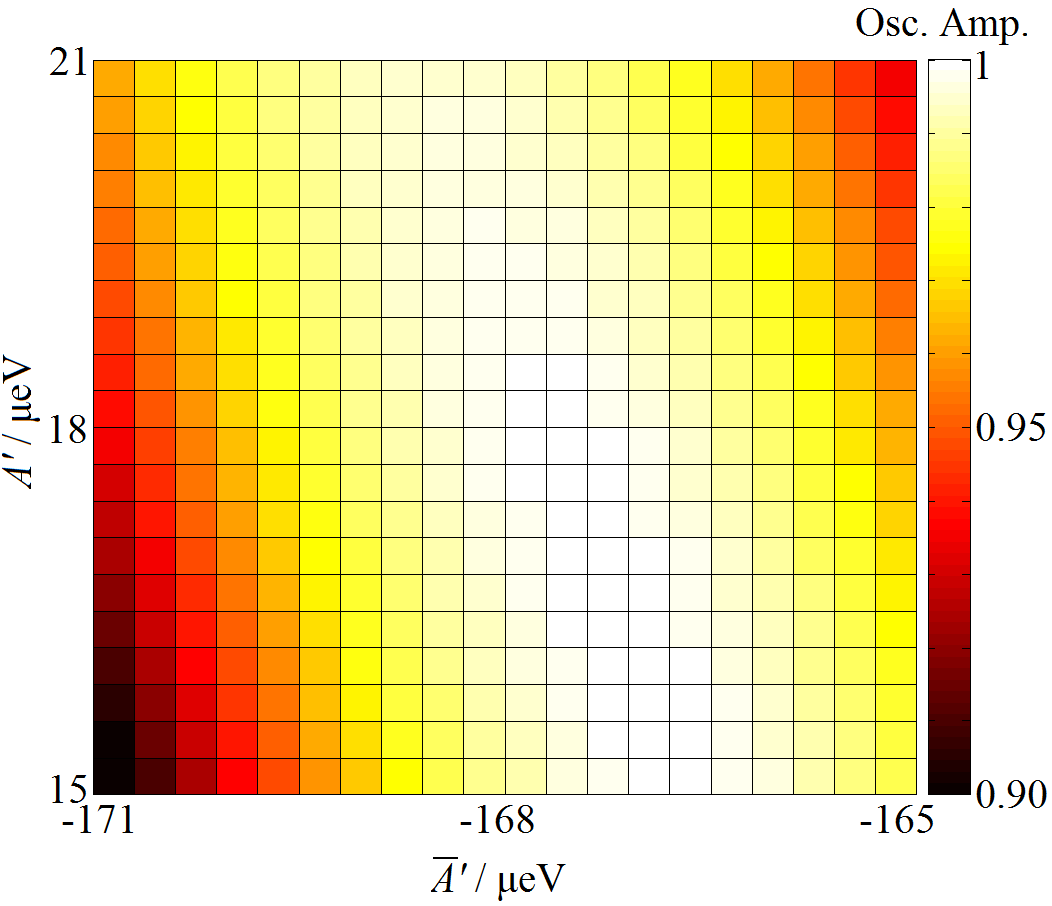}
    \caption{\label{fig:nmrDetail} A detailed plot of the highest amplitude region in Fig. \ref{fig:nmr}.}
    \end{figure}

    Having prepared the qubit in the state $\psi_0(x)$, we will now demonstrate how to perform a $\sigma_x$ rotation through an angle $\vartheta$ on the Bloch sphere.  When the response time of the quantum system is negligible, ie. $\tau = 0$, this rotation is achieved by switching the detuning to zero.  For $\varepsilon=0$, the Hamiltonian in Eq. \ref{eq:Heff} takes the form
    \begin{equation}
    \hat{H}_\mathrm{eff} = - \frac{1}{2} \,\Delta \,\sigma_{x}
    \end{equation}
    so that the rotation in Eq. \ref{eq:arbitraryRotation} becomes
    \begin{equation}
    R_{\vec{n}} (\vartheta(t))=R_{\vec{x}} (\vartheta(t)) = \exp \left( \frac{-\Delta \,\sigma_{x}}{2 i\hbar} t \right).
    \end{equation}
    Once the qubit has rotated by the desired angle, the detuning is returned to its original value.  With finite rise times, this process translates to a trapezoidal pulse shape similar to the one used for the state preparation in Sec.~\ref{sec:state_preparation}.

    Significantly, the trapezoidal pulse is not able to produce a pure $\sigma_x$ rotation.  To show this, we initiated the qubit in the $\psi_0(x)$ state and swept $A$ in the range described in Sec.~\ref{subsec:optimalqubit} and $t_{p}$ from 0 to 1000 ps (enough for all the rotations to make at least one full circle on the Bloch sphere).  After choosing the values for which the oscillation amplitude, defined as the difference between the maximum and the minimum probabilities of finding the qubit in the other state $\psi_1(x)$, was the highest, we used these optimal pulse parameters on the set of three mutually orthogonal states from Sec. \ref{sec:qubitcontrol} to map the pulse to the rotation:
    \begin{equation}
        \label{eq:sigmaxyStandard}
          R_{\vec{n}} (\vartheta(t))=R_{\vec{x}}(\kappa (t - 2 \tau)) R_{\vec{n}_{\mathrm{trans}}}(\vartheta_{\mathrm{trans}}),
    \end{equation}
    where $\vec{n}_{\mathrm{trans}}$ is the axis of a transient rotation due to the finite rise and, for the tunnelling energy and pulse rise times we have chosen, $\kappa = 0.029$ and $\vartheta_{\mathrm{trans}} = \pi - 0.10$.  Importantly, the transient rotation axis $\vec{n}_{\mathrm{trans}}$ is not aligned with the $\sigma_z$ axis so these rotations do not commute and cannot be translated into a pure $\sigma_x$ rotation.  With the trapezoidal pulse, we were therefore unable to find any values of $A$ and $t_p$ for which the qubit experienced a pure $\sigma_x$ rotation on the Bloch sphere.

    The solution, inspired by the spin-echo technique~\cite{hahn50}, was to use the modified pulse shown in Fig.~\ref{fig:pulsing} (b).  The counter-detuning, of amplitude $\bar{A}'$, applied at the beginning and end of the pulse was introduced to correct the transient rotation generated by the finite rise time.  We swept the parameters $A', \bar{A}'$ and measured the oscillation amplitude as we vary $t_p$.  The resulting amplitudes are shown in Fig.~\ref{fig:nmr}.  For a high fidelity $\sigma_x$ rotation, the state should oscillate between $\psi_0(x)$ and $\psi_1(x)$, so that the amplitude of the oscillation should equal 1.  We found the region with the highest oscillation amplitude and performed more detailed sweeps around it, as shown in Fig. \ref{fig:nmrDetail}. Testing the results on the same set of three mutually orthogonal states, we were able to generate a qubit rotation of the form
    \begin{equation}
        \label{eq:sigmaxNMR}
        R_{\vec{n}} (\vartheta(t))=R_{\vec{x}} (\vartheta_0 + \kappa' (t - 4 \tau)) R_{\vec{z}} (\pi),
    \end{equation}
    where $\vartheta_0 = -1.416$ and $\kappa' = -0.031$ for $\bar{A}' = -167.4\mu$eV and $A' = 16.5\mu$eV.  This represents a $\sigma_x$ rotation of angle $\vartheta_0 + \kappa' (t - 4 \tau)$ together with an additional $\pi$ rotation around the $\sigma_z$ axis.  This additional rotation can be corrected either by performing the qubit operation twice or by relabelling the qubit states $\psi_0(x) \to \psi_0(x), \psi_1(x) \to -\psi_1(x)$.

    Establishing this type of rotation for an experimental DQD is straightforward.  Using the technique for readout of the optimal qubit described in Sec. \ref{sec:readout} we can relate the measured probability with the qubit coefficients, and therefore identify values of $A'$ and $\bar{A'}$ that give the largest oscillation amplitude.  Once the values of $A'$ and $\bar{A}'$ are determined, the parameters $\vartheta$ and $\kappa'$ can be found by measuring the resulting probability for the shortest pulse (for $t_p = 4 \tau$) and the oscillation frequency, respectively.

\subsection{$\sigma_z$ rotations}

    We applied a similar procedure to find optimal pulse parameters for performing a $\sigma_z$ rotation on the Bloch sphere.  Again, the trapezoidal pulse was not able to generate a pure $\sigma_z$ rotation.  To show this, we initialised the qubit in the $\psi_0(x)$ state and performed a parameter sweep for a trapezoidal pulse with a maximum possible detuning of $A_{\mathrm{max}} = 200\mu$eV, i.e. within the range given in Fig.~\ref{fig:LocalisationOverlap}.  Again, transient rotations dominated these operations and it was not possible to obtain a rotation where the oscillation amplitude between the qubits was not less than $25\%$ (a high fidelity $\sigma_z$ rotation only introduces a phase difference, thus there should be no change in the amplitude).  Therefore, we again tried the modified pulse from Sec.~\ref{subsec:sigmax} (Fig.~\ref{fig:pulsing} (b)) and used the sweep from Fig.~\ref{fig:nmr}, this time focusing on the regions with the lowest amplitude.  As a result, we were able to successfully find a qubit operation where the oscillation amplitude was small (less than 0.003).  Once again, we mapped out the operation using the set of three mutually orthogonal eigenstates and found a rotation around the $\sigma_z$ axis of the Bloch sphere with the form:
    \begin{equation}
        \label{eq:sigmazNMR}
        R(t) = R_{\vec{z}} (\vartheta_0' + \kappa'' (t - 4 \tau)),
    \end{equation}
    For our system, $\vartheta_0' = 2.658$ and $\kappa'' = 0.359$ with pulse detunings $\bar{A} = -181.2\mu$eV and $A = 177.0\mu$eV.  It is worth noting that, as opposed to standard Ramsey interferometry-type experiments, this $\sigma_z$ rotation is achieved using a single pulse without the need for additional $\pi$ pulses around the perpendicular ($\sigma_x$) axis \cite{carr54}.

    Similarly to the approach discussed in Sec. \ref{subsec:sigmax}, the values for $A'$ and $\bar{A}'$ can be found by relating the measured probability with the qubit coefficients, as described in Sec. \ref{sec:readout}.

\section{Readout}
    \label{sec:readout}

    In experimental setups, it is the probability of finding the electron in one of the dots which is measured rather than the qubit superposition weighting coefficients.  We can write both qubits defined in Sec. \ref{subsec:optimalqubit} in terms of their right and left dots parts:
    \begin{eqnarray}
        \label{eq:LRzero}
        \psi_0(x) = f_{0L}(x) + f_{0R}(x) \\
        \label{eq:LRone}
        \psi_1(x) = f_{1L}(x) + f_{1R}(x)
    \end{eqnarray}
    Because the qubits $\psi_0(x)$ and $\psi_1(x)$ are orthogonal and also there is no overlap between any left and right dots wavefunction parts, we have:
    \begin{multline}
    \label{eq:readoutA}0 = \int \psi^*_0(x) \psi_1(x) dx = \\
    \int f^*_{0L}(x) f_{1L}(x) dx + \int f^*_{0L}(x) f_{1R}(x) dx \,\,+ \\
    \int f^*_{0R}(x) f_{1L}(x) dx + \int f^*_{0R}(x) f_{1R}(x) dx = \\
    \int f^*_{0L}(x) f_{1L}(x) dx + \int f^*_{0R}(x) f_{1R}(x) dx.
    \end{multline}
    The qubits are mirror images of each other, such that $\psi_0(x)$ has the same spatial distribution in the left (right) dot as $\psi_1(x)$ has in the right (left) one.  Therefore eq. \ref{eq:readoutA} implies that:
    \begin{equation}\label{eq:readoutB}
    \int f^*_{0L}(x) f_{1L}(x) dx = \int f^*_{0R}(x) f_{1R}(x) dx = 0.
    \end{equation}
    Any arbitrary state can be written as a linear combination of the two qubits right and left dot components
    \begin{multline}\label{eq:readoutC}
    \psi(x) = \alpha \psi_0(x) + \beta \psi_1(x) = \\
    \alpha \Big(f_{0L}(x) + f_{0R}(x)\Big) + \beta \Big(f_{1L}(x) + f_{1R}(x)\Big),
    \end{multline}
    The probability $P_{R}$ of finding the particle in the right dot is then:
    \begin{multline}
    P_{R} = \int_0^\infty \psi^*(x) \psi(x) dx = \\
    \int_0^\infty \Big( \alpha^* f^*_{0R}(x) + \beta^* f^*_{1R}(x) \Big)\,\Big( \alpha f_{0R}(x) + \beta f_{1R}(x) \Big) dx.
    \end{multline}
    Using eq. \ref{eq:readoutB}, this reduces to:
    \begin{multline}\label{eq:readoutE}
    P_{R} = \\
    |\alpha|^{2} \!\! \int_0^\infty f^*_{0R}(x) f_{0R}(x) dx \,+\, |\beta|^{2} \!\! \int_0^\infty f^*_{1R}(x) f_{1R}(x) dx \\
    \equiv |\alpha|^{2} \! P_0 \,+\, |\beta|^{2} \! P_1,
    \end{multline}
    where the integrals $P_0$ and $P_1$ can be obtained experimentally by setting the qubit in the $\psi_0(x)$ or $\psi_1(x)$ state, respectively, and measuring the probability of finding it in the right dot.
    Combining eq. \ref{eq:readoutE} with the normalisation condition for $\psi(x)$:
    \begin{equation}\label{eq:readoutF}
    |\alpha|^{2} + |\beta|^{2} = 1
    \end{equation}
    we obtain an equation relating $|\beta|$ to the probability $P_{R}$ of finding the particle in the right dot:
    \begin{equation}\label{eq:readoutG}
    |\beta|^{2} = \frac{P_{R} - P_0}{P_1 - P_0},
    \end{equation}
    together with a similar expression for $|\alpha|$.  As a result, the states $\psi_0(x)$ and $\psi_1(x)$ defined in Sec.~\ref{sec:methods} are not only optimal in terms of overlap with other states, but they are also the most suitable states for charge-detection readout as the probability weight of the particle being in each dot can be directly related to the modulus-squared of the weighting coefficients of the qubit superposition $\modulus{\alpha}^2$ and $\modulus{\beta}^2$.

\section{Discussion}
 \label{sec:discussion}
    The rotations given in Eqs.~\ref{eq:sigmaxyStandard}--\ref{eq:sigmazNMR} needed to be calculated using three mutually-orthogonal eigenstates.  As discussed in Sec.~\ref{sec:readout}, the state of a charge qubit can be read-out by performing measurements of the probability amplitude for finding the electron in one of the quantum dots.  However, in order to find out which rotation has been performed on a qubit for a given pulse sequence requires tomography to be performed on more than one initial state.

    Taking the $\sigma_x$ rotation studied in Sec.~\ref{subsec:sigmax} as an example, if we initialise the qubit in the $\psi_0(x)$ state, then an ideal rotation will result in the probability amplitude for finding the electron in the right dot oscillating between its smallest and largest values as the length of the pulse is increased.  However, for a pulse with finite rise times, a perfect oscillation of the electron from the left to the right dot does not guarantee that the qubit operation $R_{\vec{n}} (\vartheta)$ is a rotation around the $x$ axis, ie. $\vec{n} \neq \vec{x}$.

    The finite rise time induces a transient rotation which can only be resolved by mapping the rotation completely using three mutually-orthogonal eigenstates.  If varying a pulse parameter results in the path of the final state tracing out a rotation on the Bloch sphere, then it does not mean that this is the rotation which the qubit has experienced.

\section{Conclusions}
\label{sec:conclusion}
    In this paper, we have described the optimal charge qubits for a double-quantum dot system and presented pulse sequences for state preparation and arbitrary qubit rotation where the experimental response suffers from finite rise times.  We showed that, due to hybridisation of the eigenstates in a double-dot system, the spatial wave function of the two lowest energy eigenstates cannot be confined exclusively to the left and the right dot.  The qubits which have the greatest overlap to all other eigenstates were found to be defined in terms of the two lowest energy eigenstates when the dots were on resonance.  This allowed us to reduce our model to a two-state system.

    We showed that it is possible to prepare the qubit in such a state when it is initially in the eigenstate of a DQD with a non-zero detuning.  With trapezoidal pulses it was not possible to perform arbitrary qubit rotations when the detuning pulses were subject to finite rise times due to unwanted transient rotations.  Using a pulse inspired by spin-echo techniques, which included a counter-detuning to correct the transient rotation, $\sigma_x$ and $\sigma_z$ rotations of an arbitrary angle were possible.  We also discussed how the qubit state can be experimentally read-out and how tomography of the qubit oscillation must be performed with three mutually-orthogonal eigenstates for this DQD system.

    This work was supported by the Project for Developing Innovation Systems of the Ministry of Education, Culture, Sports, Science and Technology (MEXT), Japan.

\appendix

\section{Iteration method}\label{subsec:ischeme}
    The system is modelled using an explicit iterative scheme for the one-dimensional time-dependent Schr\"odinger equation (TDSE) with an arbitrary potential \emph{V(x,t)}:
    \begin{equation}\label{eq:TDSE}i \hbar \frac{\partial \psi(x, t)}{\partial t} = H\psi = \left[\frac{-\hbar^{2}}{2 m} \frac{\partial^{2}}{\partial x^{2}} + V(x, t)\right] \psi(x, t)\end{equation}
    where \emph{m} is the effective mass. The scheme, which is based on the finite difference method, was described in details by Maestri \emph{et al.} for two particles in one dimension \cite{maestri00} and we adapt it to a single particle. The wavefunction is evaluated on a spatially discretized grid and at successive, equally separated intervals of time $\Delta t$:
    \begin{equation}\label{eq:psi}\psi(x, t) = \psi(m \Delta x, k \Delta t) \equiv \psi^{k}_{m},\end{equation}
    with $m, k$ integer.
    The spatial part of the method is derived using Taylor expansion of the wavefunction:
    \begin{equation}\label{eq:xderiv}\frac{\partial^{2} \psi}{\partial x^{2}} \simeq \frac{\psi(x+\Delta x) - 2 \psi(x) + \psi(x-\Delta x)}{\Delta x^{2}}. \end{equation}
    Therefore, using eq. (\ref{eq:psi}) and (\ref{eq:xderiv}), the right hand side of eq. (\ref{eq:TDSE}) transforms into
    \begin{multline}\label{eq:TDSEdiscrete}
    H\psi = \Bigg[\frac{-\hbar^{2}}{2 m}\bigg( \frac{\psi_{m+1} - 2 \psi_{m} + \psi_{m-1} }{\Delta x^{2}} \bigg)\, +  \,V_{m}\Bigg] \psi_{m}.
    \end{multline}
    The derivative on the left hand side of eq. (\ref{eq:TDSE}) is calculated by writing the exact solution of TDSE and then taking the difference between the $(k\!+\!1)^{th}$ and $(k\!-\!1)^{th}$ time steps, as suggested by Askar and Cakmak \cite{askar78}:
    \begin{equation}\label{eq:psinext}\psi^{k+1}_{m} = e^{-i \Delta t H / \hbar} \psi^{k}_{m} \simeq \left(1 - \frac{i \Delta t H}{\hbar}\right) \psi^{k}_{m},\end{equation}
    \begin{multline}\label{eq:psidiff}
    \psi^{k+1}_{m}\, - \,\psi^{k-1}_{m} = (e^{-i \Delta t H / \hbar} - e^{i \Delta t H / \hbar}) \psi^{k}_{m} \\ \simeq -2 \left(\frac{i \Delta t H}{\hbar}\right) \psi^{k}_{m}.
    \end{multline}
    To improve the accuracy, we follow Visscher's staggered-time method \cite{visscher91} and write the wavevector in terms of its real and imaginary parts: $\psi^{k}_{m} = u^{k}_{m} + i v^{k}_{m}$. After inserting the Hamiltonian from eq. (\ref{eq:TDSEdiscrete}) into eq. (\ref{eq:psidiff}) and rearranging the terms, we obtain a pair of simultaneous equations, which are iterated over time:
    \begin{equation}\begin{aligned}\label{eq:uv}
    u^{k+1}_{m} &= u^{k-1}_{m} + \bigg[ \Big( 2a_{x}+b\,V^{k}_{m} \Big) v^{k}_{m} - a_{x} (v^{k}_{m+1}\!+\!v^{k}_{m-1}) \bigg],\\
    v^{k+1}_{m} &= v^{k-1}_{m} - \bigg[ \Big( 2a_{x}+b\,V^{k}_{m} \Big) u^{k}_{m} - a_{x} (u^{k}_{m+1}\!+\!u^{k}_{m-1}) \bigg],
    \end{aligned}\end{equation}
    where $a_{x}=\frac{\hbar \Delta t}{m \Delta x^{2}}$ and $b=\frac{2 \Delta t}{\hbar}$. Also, the real and imaginary parts are calculated at slightly shifted times: $u^{k} \equiv u(t),\; v^{k} \equiv v(t+\Delta t/2).$

    The method above is stable as long as the following criterion is satisfied:
    \begin{equation}
    \Delta t \leq \frac{\hbar}{E_{max}},
    \end{equation}
    with $E_{max}$ being the largest eigenvalue of the discretised Hamiltonian \cite{leforestier91}. Furthermore, small errors due to finite computational accuracy do not accumulate with iterations and the total electron probability $\sum_{\mathrm{all\,} m} |\psi^{k}_{m}|^2$ is preserved over time, showing no significant deviations from unity.

\section{GPU acceleration}\label{subsec:gpu}
    The scheme described in App.~\ref{subsec:ischeme} is computationally demanding. However, the set of defining equations (\ref{eq:uv}) can be parallelised in a straightforward manner:  the next value of an element at position $(m)$ depends only on its value two steps before and the previous values of its neighbours at positions $(m-1)$ and $(m+1)$. Therefore, each vector element can be calculated separately and the whole procedure can be divided into a set of $N_{x}$ simultaneous steps \cite{owen12}.\par
    To further employ parallelism, calculations were transferred to GPUs, as graphics cards are excellent at accelerating processes involving large numbers of smaller jobs evaluated concurrently. In this case, these separate tasks were the calculations for the next vector elements at each node. Our code was written using OpenCL, which provides a framework for writing programs that run on arbitrary GPUs \cite{opencl14}. Comparison tests performed with a single computer showed a performance improvement of two orders of magnitude using GPU accelerated code over CPU code.


\bibliography{paper}

\providecommand{\noopsort}[1]{}\providecommand{\singleletter}[1]{#1}%
\begin{thebibliography}{36}%
\makeatletter
\providecommand \@ifxundefined [1]{%
 \@ifx{#1\undefined}
}%
\providecommand \@ifnum [1]{%
 \ifnum #1\expandafter \@firstoftwo
 \else \expandafter \@secondoftwo
 \fi
}%
\providecommand \@ifx [1]{%
 \ifx #1\expandafter \@firstoftwo
 \else \expandafter \@secondoftwo
 \fi
}%
\providecommand \natexlab [1]{#1}%
\providecommand \enquote  [1]{``#1''}%
\providecommand \bibnamefont  [1]{#1}%
\providecommand \bibfnamefont [1]{#1}%
\providecommand \citenamefont [1]{#1}%
\providecommand \href@noop [0]{\@secondoftwo}%
\providecommand \href [0]{\begingroup \@sanitize@url \@href}%
\providecommand \@href[1]{\@@startlink{#1}\@@href}%
\providecommand \@@href[1]{\endgroup#1\@@endlink}%
\providecommand \@sanitize@url [0]{\catcode `\\12\catcode `\$12\catcode
  `\&12\catcode `\#12\catcode `\^12\catcode `\_12\catcode `\%12\relax}%
\providecommand \@@startlink[1]{}%
\providecommand \@@endlink[0]{}%
\providecommand \url  [0]{\begingroup\@sanitize@url \@url }%
\providecommand \@url [1]{\endgroup\@href {#1}{\urlprefix }}%
\providecommand \urlprefix  [0]{URL }%
\providecommand \Eprint [0]{\href }%
\providecommand \doibase [0]{http://dx.doi.org/}%
\providecommand \selectlanguage [0]{\@gobble}%
\providecommand \bibinfo  [0]{\@secondoftwo}%
\providecommand \bibfield  [0]{\@secondoftwo}%
\providecommand \translation [1]{[#1]}%
\providecommand \BibitemOpen [0]{}%
\providecommand \bibitemStop [0]{}%
\providecommand \bibitemNoStop [0]{.\EOS\space}%
\providecommand \EOS [0]{\spacefactor3000\relax}%
\providecommand \BibitemShut  [1]{\csname bibitem#1\endcsname}%
\let\auto@bib@innerbib\@empty
\bibitem [{\citenamefont {Hayashi}\ \emph {et~al.}(2003)\citenamefont
  {Hayashi}, \citenamefont {Fujisawa}, \citenamefont {Cheong}, \citenamefont
  {Jeong},\ and\ \citenamefont {Hirayama}}]{hayashi03}%
  \BibitemOpen
  \bibfield  {author} {\bibinfo {author} {\bibfnamefont {T.}~\bibnamefont
  {Hayashi}}, \bibinfo {author} {\bibfnamefont {T.}~\bibnamefont {Fujisawa}},
  \bibinfo {author} {\bibfnamefont {H.~D.}\ \bibnamefont {Cheong}}, \bibinfo
  {author} {\bibfnamefont {Y.~H.}\ \bibnamefont {Jeong}}, \ and\ \bibinfo
  {author} {\bibfnamefont {Y.}~\bibnamefont {Hirayama}},\ }\href@noop {}
  {\bibfield  {journal} {\bibinfo  {journal} {Phys. Rev. Lett.}\ }\textbf
  {\bibinfo {volume} {91}},\ \bibinfo {pages} {226804} (\bibinfo {year}
  {2003})}\BibitemShut {NoStop}%
\bibitem [{\citenamefont {Fujisawa}\ \emph {et~al.}(2006)\citenamefont
  {Fujisawa}, \citenamefont {Hayashi},\ and\ \citenamefont
  {Sasaki}}]{fujisawa06}%
  \BibitemOpen
  \bibfield  {author} {\bibinfo {author} {\bibfnamefont {T.}~\bibnamefont
  {Fujisawa}}, \bibinfo {author} {\bibfnamefont {T.}~\bibnamefont {Hayashi}}, \
  and\ \bibinfo {author} {\bibfnamefont {S.}~\bibnamefont {Sasaki}},\
  }\href@noop {} {\bibfield  {journal} {\bibinfo  {journal} {Rep. Prog. Phys.}\
  }\textbf {\bibinfo {volume} {69}},\ \bibinfo {pages} {759} (\bibinfo {year}
  {2006})}\BibitemShut {NoStop}%
\bibitem [{\citenamefont {Petta}\ \emph {et~al.}(2005)\citenamefont {Petta},
  \citenamefont {Johnson}, \citenamefont {Taylor}, \citenamefont {Laird},
  \citenamefont {Yacoby}, \citenamefont {Lukin}, \citenamefont {Marcus},
  \citenamefont {Hanson},\ and\ \citenamefont {Gossard}}]{petta05}%
  \BibitemOpen
  \bibfield  {author} {\bibinfo {author} {\bibfnamefont {J.~R.}\ \bibnamefont
  {Petta}}, \bibinfo {author} {\bibfnamefont {A.~C.}\ \bibnamefont {Johnson}},
  \bibinfo {author} {\bibfnamefont {J.~M.}\ \bibnamefont {Taylor}}, \bibinfo
  {author} {\bibfnamefont {E.~A.}\ \bibnamefont {Laird}}, \bibinfo {author}
  {\bibfnamefont {A.}~\bibnamefont {Yacoby}}, \bibinfo {author} {\bibfnamefont
  {M.~D.}\ \bibnamefont {Lukin}}, \bibinfo {author} {\bibfnamefont {C.~M.}\
  \bibnamefont {Marcus}}, \bibinfo {author} {\bibfnamefont {M.~P.}\
  \bibnamefont {Hanson}}, \ and\ \bibinfo {author} {\bibfnamefont {A.~C.}\
  \bibnamefont {Gossard}},\ }\href@noop {} {\bibfield  {journal} {\bibinfo
  {journal} {Science}\ }\textbf {\bibinfo {volume} {309}},\ \bibinfo {pages}
  {2180} (\bibinfo {year} {2005})}\BibitemShut {NoStop}%
\bibitem [{\citenamefont {Petersson}\ \emph {et~al.}(2010)\citenamefont
  {Petersson}, \citenamefont {Petta}, \citenamefont {Lu},\ and\ \citenamefont
  {Gossard}}]{petersson10}%
  \BibitemOpen
  \bibfield  {author} {\bibinfo {author} {\bibfnamefont {K.~D.}\ \bibnamefont
  {Petersson}}, \bibinfo {author} {\bibfnamefont {J.~R.}\ \bibnamefont
  {Petta}}, \bibinfo {author} {\bibfnamefont {H.}~\bibnamefont {Lu}}, \ and\
  \bibinfo {author} {\bibfnamefont {A.~C.}\ \bibnamefont {Gossard}},\
  }\href@noop {} {\bibfield  {journal} {\bibinfo  {journal} {Phys. Rev. Lett.}\
  }\textbf {\bibinfo {volume} {105}},\ \bibinfo {pages} {246804} (\bibinfo
  {year} {2010})}\BibitemShut {NoStop}%
\bibitem [{\citenamefont {Veldhorst}\ \emph {et~al.}(2014)\citenamefont
  {Veldhorst}, \citenamefont {Hwang}, \citenamefont {Yang}, \citenamefont
  {Leenstra}, \citenamefont {de~Ronde}, \citenamefont {Dehollain},
  \citenamefont {Muhonen}, \citenamefont {Hudson}, \citenamefont {Itoh},
  \citenamefont {Morello},\ and\ \citenamefont {Dzurak}}]{veldhorst14}%
  \BibitemOpen
  \bibfield  {author} {\bibinfo {author} {\bibfnamefont {M.}~\bibnamefont
  {Veldhorst}}, \bibinfo {author} {\bibfnamefont {J.~C.~C.}\ \bibnamefont
  {Hwang}}, \bibinfo {author} {\bibfnamefont {C.~H.}\ \bibnamefont {Yang}},
  \bibinfo {author} {\bibfnamefont {A.~W.}\ \bibnamefont {Leenstra}}, \bibinfo
  {author} {\bibfnamefont {B.}~\bibnamefont {de~Ronde}}, \bibinfo {author}
  {\bibfnamefont {J.~P.}\ \bibnamefont {Dehollain}}, \bibinfo {author}
  {\bibfnamefont {J.~T.}\ \bibnamefont {Muhonen}}, \bibinfo {author}
  {\bibfnamefont {F.~E.}\ \bibnamefont {Hudson}}, \bibinfo {author}
  {\bibfnamefont {K.~M.}\ \bibnamefont {Itoh}}, \bibinfo {author}
  {\bibfnamefont {A.}~\bibnamefont {Morello}}, \ and\ \bibinfo {author}
  {\bibfnamefont {A.~S.}\ \bibnamefont {Dzurak}},\ }\href@noop {} {\bibfield
  {journal} {\bibinfo  {journal} {Nature Nanotech.}\ }\textbf {\bibinfo
  {volume} {9}},\ \bibinfo {pages} {981} (\bibinfo {year} {2014})}\BibitemShut
  {NoStop}%
\bibitem [{\citenamefont {Lloyd}(1993)}]{lloyd93}%
  \BibitemOpen
  \bibfield  {author} {\bibinfo {author} {\bibfnamefont {S.}~\bibnamefont
  {Lloyd}},\ }\href@noop {} {\bibfield  {journal} {\bibinfo  {journal}
  {Science}\ }\textbf {\bibinfo {volume} {261}},\ \bibinfo {pages} {1569}
  (\bibinfo {year} {1993})}\BibitemShut {NoStop}%
\bibitem [{\citenamefont {DiVincenzo}(2000)}]{divincenzo00}%
  \BibitemOpen
  \bibfield  {author} {\bibinfo {author} {\bibfnamefont {D.~P.}\ \bibnamefont
  {DiVincenzo}},\ }\href@noop {} {\bibfield  {journal} {\bibinfo  {journal}
  {Fortschritte der Physik}\ }\textbf {\bibinfo {volume} {48}},\ \bibinfo
  {pages} {771} (\bibinfo {year} {2000})}\BibitemShut {NoStop}%
\bibitem [{\citenamefont {Dovzhenko}\ \emph {et~al.}(2011)\citenamefont
  {Dovzhenko}, \citenamefont {Stehlik}, \citenamefont {Petersson},
  \citenamefont {Petta}, \citenamefont {Lu},\ and\ \citenamefont
  {Gossard}}]{dovzhenko11}%
  \BibitemOpen
  \bibfield  {author} {\bibinfo {author} {\bibfnamefont {Y.}~\bibnamefont
  {Dovzhenko}}, \bibinfo {author} {\bibfnamefont {J.}~\bibnamefont {Stehlik}},
  \bibinfo {author} {\bibfnamefont {K.~D.}\ \bibnamefont {Petersson}}, \bibinfo
  {author} {\bibfnamefont {J.~R.}\ \bibnamefont {Petta}}, \bibinfo {author}
  {\bibfnamefont {H.}~\bibnamefont {Lu}}, \ and\ \bibinfo {author}
  {\bibfnamefont {A.~C.}\ \bibnamefont {Gossard}},\ }\href@noop {} {\bibfield
  {journal} {\bibinfo  {journal} {Phys. Rev. B}\ }\textbf {\bibinfo {volume}
  {84}},\ \bibinfo {pages} {161302} (\bibinfo {year} {2011})}\BibitemShut
  {NoStop}%
\bibitem [{\citenamefont {Kataoka}\ \emph {et~al.}(2009)\citenamefont
  {Kataoka}, \citenamefont {Astley}, \citenamefont {Thorn}, \citenamefont {Oi},
  \citenamefont {Barnes}, \citenamefont {Ford}, \citenamefont {Anderson},
  \citenamefont {Jones}, \citenamefont {Farrer}, \citenamefont {Ritchie},\ and\
  \citenamefont {Pepper}}]{kataoka09}%
  \BibitemOpen
  \bibfield  {author} {\bibinfo {author} {\bibfnamefont {M.}~\bibnamefont
  {Kataoka}}, \bibinfo {author} {\bibfnamefont {M.~R.}\ \bibnamefont {Astley}},
  \bibinfo {author} {\bibfnamefont {A.~L.}\ \bibnamefont {Thorn}}, \bibinfo
  {author} {\bibfnamefont {D.~K.~L.}\ \bibnamefont {Oi}}, \bibinfo {author}
  {\bibfnamefont {C.~H.~W.}\ \bibnamefont {Barnes}}, \bibinfo {author}
  {\bibfnamefont {C.~J.~B.}\ \bibnamefont {Ford}}, \bibinfo {author}
  {\bibfnamefont {D.}~\bibnamefont {Anderson}}, \bibinfo {author}
  {\bibfnamefont {G.~A.~C.}\ \bibnamefont {Jones}}, \bibinfo {author}
  {\bibfnamefont {I.}~\bibnamefont {Farrer}}, \bibinfo {author} {\bibfnamefont
  {D.~A.}\ \bibnamefont {Ritchie}}, \ and\ \bibinfo {author} {\bibfnamefont
  {M.}~\bibnamefont {Pepper}},\ }\href@noop {} {\bibfield  {journal} {\bibinfo
  {journal} {Phys. Rev. Lett.}\ }\textbf {\bibinfo {volume} {102}},\ \bibinfo
  {pages} {156801} (\bibinfo {year} {2009})}\BibitemShut {NoStop}%
\bibitem [{\citenamefont {Rossi}\ \emph {et~al.}(2010)\citenamefont {Rossi},
  \citenamefont {Ferrus}, \citenamefont {Podd},\ and\ \citenamefont
  {Williams}}]{rossi10}%
  \BibitemOpen
  \bibfield  {author} {\bibinfo {author} {\bibfnamefont {A.}~\bibnamefont
  {Rossi}}, \bibinfo {author} {\bibfnamefont {T.}~\bibnamefont {Ferrus}},
  \bibinfo {author} {\bibfnamefont {G.~J.}\ \bibnamefont {Podd}}, \ and\
  \bibinfo {author} {\bibfnamefont {D.~A.}\ \bibnamefont {Williams}},\
  }\href@noop {} {\bibfield  {journal} {\bibinfo  {journal} {Appl. Phys.
  Lett.}\ }\textbf {\bibinfo {volume} {97}},\ \bibinfo {pages} {223506}
  (\bibinfo {year} {2010})}\BibitemShut {NoStop}%
\bibitem [{\citenamefont {Owen}\ and\ \citenamefont {Barnes}(2015)}]{owen15}%
  \BibitemOpen
  \bibfield  {author} {\bibinfo {author} {\bibfnamefont {E.~T.}\ \bibnamefont
  {Owen}}\ and\ \bibinfo {author} {\bibfnamefont {C.~H.~W.}\ \bibnamefont
  {Barnes}},\ }\href@noop {} {\  (\bibinfo {year} {2015})},\ \Eprint
  {http://arxiv.org/abs/1509.02457} {arXiv:1509.02457} \BibitemShut {NoStop}%
\bibitem [{\citenamefont {Stopa}(1996)}]{stopa96}%
  \BibitemOpen
  \bibfield  {author} {\bibinfo {author} {\bibfnamefont {M.}~\bibnamefont
  {Stopa}},\ }\href@noop {} {\bibfield  {journal} {\bibinfo  {journal} {Phys.
  Rev. B}\ }\textbf {\bibinfo {volume} {54}},\ \bibinfo {pages} {13767}
  (\bibinfo {year} {1996})}\BibitemShut {NoStop}%
\bibitem [{\citenamefont {Ferrus}\ \emph {et~al.}(2011)\citenamefont {Ferrus},
  \citenamefont {Rossi}, \citenamefont {Tanner}, \citenamefont {Podd},
  \citenamefont {Chapman},\ and\ \citenamefont {Williams}}]{ferrus11}%
  \BibitemOpen
  \bibfield  {author} {\bibinfo {author} {\bibfnamefont {T.}~\bibnamefont
  {Ferrus}}, \bibinfo {author} {\bibfnamefont {A.}~\bibnamefont {Rossi}},
  \bibinfo {author} {\bibfnamefont {M.}~\bibnamefont {Tanner}}, \bibinfo
  {author} {\bibfnamefont {G.}~\bibnamefont {Podd}}, \bibinfo {author}
  {\bibfnamefont {P.}~\bibnamefont {Chapman}}, \ and\ \bibinfo {author}
  {\bibfnamefont {D.~A.}\ \bibnamefont {Williams}},\ }\href@noop {} {\bibfield
  {journal} {\bibinfo  {journal} {New J. Phys.}\ }\textbf {\bibinfo {volume}
  {13}},\ \bibinfo {pages} {103012} (\bibinfo {year} {2011})}\BibitemShut
  {NoStop}%
\bibitem [{\citenamefont {D.~Hisamoto}\ and\ \citenamefont
  {Takeda}(1990)}]{hisamoto90}%
  \BibitemOpen
  \bibfield  {author} {\bibinfo {author} {\bibfnamefont {Y.~K.}\ \bibnamefont
  {D.~Hisamoto}, \bibfnamefont {T.~Kaga}}\ and\ \bibinfo {author}
  {\bibfnamefont {E.}~\bibnamefont {Takeda}},\ }\href@noop {} {\bibfield
  {journal} {\bibinfo  {journal} {IEEE Electron Device Letters}\ }\textbf
  {\bibinfo {volume} {11}},\ \bibinfo {pages} {36} (\bibinfo {year}
  {1990})}\BibitemShut {NoStop}%
\bibitem [{\citenamefont {Voisin}\ \emph {et~al.}(2014)\citenamefont {Voisin},
  \citenamefont {Nguyen}, \citenamefont {Renard}, \citenamefont {Jehl},
  \citenamefont {Barraud}, \citenamefont {Triozon}, \citenamefont {Vinet},
  \citenamefont {Duchemin}, \citenamefont {Niquet}, \citenamefont
  {de~Franceschi},\ and\ \citenamefont {Sanquer}}]{voisin14}%
  \BibitemOpen
  \bibfield  {author} {\bibinfo {author} {\bibfnamefont {B.}~\bibnamefont
  {Voisin}}, \bibinfo {author} {\bibfnamefont {V.-H.}\ \bibnamefont {Nguyen}},
  \bibinfo {author} {\bibfnamefont {J.}~\bibnamefont {Renard}}, \bibinfo
  {author} {\bibfnamefont {X.}~\bibnamefont {Jehl}}, \bibinfo {author}
  {\bibfnamefont {S.}~\bibnamefont {Barraud}}, \bibinfo {author} {\bibfnamefont
  {F.}~\bibnamefont {Triozon}}, \bibinfo {author} {\bibfnamefont
  {M.}~\bibnamefont {Vinet}}, \bibinfo {author} {\bibfnamefont
  {I.}~\bibnamefont {Duchemin}}, \bibinfo {author} {\bibfnamefont {Y.-M.}\
  \bibnamefont {Niquet}}, \bibinfo {author} {\bibfnamefont {S.}~\bibnamefont
  {de~Franceschi}}, \ and\ \bibinfo {author} {\bibfnamefont {M.}~\bibnamefont
  {Sanquer}},\ }\href@noop {} {\bibfield  {journal} {\bibinfo  {journal} {Nano
  Lett.}\ }\textbf {\bibinfo {volume} {14}},\ \bibinfo {pages} {2094} (\bibinfo
  {year} {2014})}\BibitemShut {NoStop}%
\bibitem [{\citenamefont {van Wees}\ \emph {et~al.}(1988)\citenamefont {van
  Wees}, \citenamefont {van Houten}, \citenamefont {Beenakker}, \citenamefont
  {Williamson}, \citenamefont {Kouwenhoven}, \citenamefont {van~der Marel},\
  and\ \citenamefont {Foxon}}]{vanwees88}%
  \BibitemOpen
  \bibfield  {author} {\bibinfo {author} {\bibfnamefont {B.~J.}\ \bibnamefont
  {van Wees}}, \bibinfo {author} {\bibfnamefont {H.}~\bibnamefont {van
  Houten}}, \bibinfo {author} {\bibfnamefont {C.~W.~J.}\ \bibnamefont
  {Beenakker}}, \bibinfo {author} {\bibfnamefont {J.~G.}\ \bibnamefont
  {Williamson}}, \bibinfo {author} {\bibfnamefont {L.~P.}\ \bibnamefont
  {Kouwenhoven}}, \bibinfo {author} {\bibfnamefont {D.}~\bibnamefont {van~der
  Marel}}, \ and\ \bibinfo {author} {\bibfnamefont {C.~T.}\ \bibnamefont
  {Foxon}},\ }\href@noop {} {\bibfield  {journal} {\bibinfo  {journal} {Phys.
  Rev. Lett.}\ }\textbf {\bibinfo {volume} {60}},\ \bibinfo {pages} {848}
  (\bibinfo {year} {1988})}\BibitemShut {NoStop}%
\bibitem [{\citenamefont {Fujisawa}\ \emph {et~al.}(2000)\citenamefont
  {Fujisawa}, \citenamefont {van~der Wiel},\ and\ \citenamefont
  {Kouwenhoven}}]{fujisawa00}%
  \BibitemOpen
  \bibfield  {author} {\bibinfo {author} {\bibfnamefont {T.}~\bibnamefont
  {Fujisawa}}, \bibinfo {author} {\bibfnamefont {W.~G.}\ \bibnamefont {van~der
  Wiel}}, \ and\ \bibinfo {author} {\bibfnamefont {L.~P.}\ \bibnamefont
  {Kouwenhoven}},\ }\href@noop {} {\bibfield  {journal} {\bibinfo  {journal}
  {Physica E}\ }\textbf {\bibinfo {volume} {73}},\ \bibinfo {pages} {413}
  (\bibinfo {year} {2000})}\BibitemShut {NoStop}%
\bibitem [{\citenamefont {Gardelis}\ \emph {et~al.}(2003)\citenamefont
  {Gardelis}, \citenamefont {Smith}, \citenamefont {Cooper}, \citenamefont
  {Ritchie}, \citenamefont {Linfield}, \citenamefont {Jin},\ and\ \citenamefont
  {Pepper}}]{gardelis03}%
  \BibitemOpen
  \bibfield  {author} {\bibinfo {author} {\bibfnamefont {S.}~\bibnamefont
  {Gardelis}}, \bibinfo {author} {\bibfnamefont {C.~G.}\ \bibnamefont {Smith}},
  \bibinfo {author} {\bibfnamefont {J.}~\bibnamefont {Cooper}}, \bibinfo
  {author} {\bibfnamefont {D.~A.}\ \bibnamefont {Ritchie}}, \bibinfo {author}
  {\bibfnamefont {E.~H.}\ \bibnamefont {Linfield}}, \bibinfo {author}
  {\bibfnamefont {Y.}~\bibnamefont {Jin}}, \ and\ \bibinfo {author}
  {\bibfnamefont {M.}~\bibnamefont {Pepper}},\ }\href@noop {} {\bibfield
  {journal} {\bibinfo  {journal} {Phys. Rev. B}\ }\textbf {\bibinfo {volume}
  {67}},\ \bibinfo {pages} {073302} (\bibinfo {year} {2003})}\BibitemShut
  {NoStop}%
\bibitem [{\citenamefont {Lim}\ \emph {et~al.}(2009)\citenamefont {Lim},
  \citenamefont {Huebl}, \citenamefont {van Beveren}, \citenamefont {Rubanov},
  \citenamefont {Spizzirri}, \citenamefont {Angus}, \citenamefont {Clark},\
  and\ \citenamefont {Dzurak}}]{lim09}%
  \BibitemOpen
  \bibfield  {author} {\bibinfo {author} {\bibfnamefont {W.~H.}\ \bibnamefont
  {Lim}}, \bibinfo {author} {\bibfnamefont {H.}~\bibnamefont {Huebl}}, \bibinfo
  {author} {\bibfnamefont {L.~H.~W.}\ \bibnamefont {van Beveren}}, \bibinfo
  {author} {\bibfnamefont {S.}~\bibnamefont {Rubanov}}, \bibinfo {author}
  {\bibfnamefont {P.~G.}\ \bibnamefont {Spizzirri}}, \bibinfo {author}
  {\bibfnamefont {S.~J.}\ \bibnamefont {Angus}}, \bibinfo {author}
  {\bibfnamefont {R.~G.}\ \bibnamefont {Clark}}, \ and\ \bibinfo {author}
  {\bibfnamefont {A.~S.}\ \bibnamefont {Dzurak}},\ }\href@noop {} {\bibfield
  {journal} {\bibinfo  {journal} {Appl. Phys. Lett.}\ }\textbf {\bibinfo
  {volume} {94}} (\bibinfo {year} {2009})}\BibitemShut {NoStop}%
\bibitem [{\citenamefont {Mason}\ \emph {et~al.}(2004)\citenamefont {Mason},
  \citenamefont {Biercuk},\ and\ \citenamefont {Marcus}}]{mason04}%
  \BibitemOpen
  \bibfield  {author} {\bibinfo {author} {\bibfnamefont {N.}~\bibnamefont
  {Mason}}, \bibinfo {author} {\bibfnamefont {M.~J.}\ \bibnamefont {Biercuk}},
  \ and\ \bibinfo {author} {\bibfnamefont {C.~M.}\ \bibnamefont {Marcus}},\
  }\href@noop {} {\bibfield  {journal} {\bibinfo  {journal} {Science}\ }\textbf
  {\bibinfo {volume} {303}},\ \bibinfo {pages} {655} (\bibinfo {year}
  {2004})}\BibitemShut {NoStop}%
\bibitem [{\citenamefont {Wei}\ \emph {et~al.}(2013)\citenamefont {Wei},
  \citenamefont {Li}, \citenamefont {Cao}, \citenamefont {Luo}, \citenamefont
  {Zheng}, \citenamefont {Tu}, \citenamefont {Xiao}, \citenamefont {Guo},
  \citenamefont {Jiang},\ and\ \citenamefont {Guo}}]{wei13}%
  \BibitemOpen
  \bibfield  {author} {\bibinfo {author} {\bibfnamefont {D.}~\bibnamefont
  {Wei}}, \bibinfo {author} {\bibfnamefont {H.-O.}\ \bibnamefont {Li}},
  \bibinfo {author} {\bibfnamefont {G.}~\bibnamefont {Cao}}, \bibinfo {author}
  {\bibfnamefont {G.}~\bibnamefont {Luo}}, \bibinfo {author} {\bibfnamefont
  {Z.-X.}\ \bibnamefont {Zheng}}, \bibinfo {author} {\bibfnamefont
  {T.}~\bibnamefont {Tu}}, \bibinfo {author} {\bibfnamefont {M.}~\bibnamefont
  {Xiao}}, \bibinfo {author} {\bibfnamefont {G.-C.}\ \bibnamefont {Guo}},
  \bibinfo {author} {\bibfnamefont {H.-W.}\ \bibnamefont {Jiang}}, \ and\
  \bibinfo {author} {\bibfnamefont {G.-P.}\ \bibnamefont {Guo}},\ }\href@noop
  {} {\bibfield  {journal} {\bibinfo  {journal} {Sci. Rep.}\ }\textbf {\bibinfo
  {volume} {3}} (\bibinfo {year} {2013})}\BibitemShut {NoStop}%
\bibitem [{\citenamefont {Askar}\ and\ \citenamefont {Cakmak}(1978)}]{askar78}%
  \BibitemOpen
  \bibfield  {author} {\bibinfo {author} {\bibfnamefont {A.}~\bibnamefont
  {Askar}}\ and\ \bibinfo {author} {\bibfnamefont {A.~S.}\ \bibnamefont
  {Cakmak}},\ }\href@noop {} {\bibfield  {journal} {\bibinfo  {journal} {J.
  Chem. Phys.}\ }\textbf {\bibinfo {volume} {68}},\ \bibinfo {pages} {2794}
  (\bibinfo {year} {1978})}\BibitemShut {NoStop}%
\bibitem [{\citenamefont {Owen}\ \emph {et~al.}(2012)\citenamefont {Owen},
  \citenamefont {Dean},\ and\ \citenamefont {Barnes}}]{owen12}%
  \BibitemOpen
  \bibfield  {author} {\bibinfo {author} {\bibfnamefont {E.~T.}\ \bibnamefont
  {Owen}}, \bibinfo {author} {\bibfnamefont {M.~C.}\ \bibnamefont {Dean}}, \
  and\ \bibinfo {author} {\bibfnamefont {C.~H.~W.}\ \bibnamefont {Barnes}},\
  }\href@noop {} {\bibfield  {journal} {\bibinfo  {journal} {Phys. Rev. A}\
  }\textbf {\bibinfo {volume} {85}},\ \bibinfo {pages} {022319} (\bibinfo
  {year} {2012})}\BibitemShut {NoStop}%
\bibitem [{\citenamefont {Fujisawa}\ \emph {et~al.}(2004)\citenamefont
  {Fujisawa}, \citenamefont {Hayashi}, \citenamefont {Cheong}, \citenamefont
  {Jeong},\ and\ \citenamefont {Hirayama}}]{fujisawa04}%
  \BibitemOpen
  \bibfield  {author} {\bibinfo {author} {\bibfnamefont {T.}~\bibnamefont
  {Fujisawa}}, \bibinfo {author} {\bibfnamefont {T.}~\bibnamefont {Hayashi}},
  \bibinfo {author} {\bibfnamefont {H.~D.}\ \bibnamefont {Cheong}}, \bibinfo
  {author} {\bibfnamefont {Y.~H.}\ \bibnamefont {Jeong}}, \ and\ \bibinfo
  {author} {\bibfnamefont {Y.}~\bibnamefont {Hirayama}},\ }\href@noop {}
  {\bibfield  {journal} {\bibinfo  {journal} {Physica E}\ }\textbf {\bibinfo
  {volume} {21}},\ \bibinfo {pages} {1046} (\bibinfo {year}
  {2004})}\BibitemShut {NoStop}%
\bibitem [{com()}]{comment1}%
  \BibitemOpen
  \href@noop {} {\bibinfo  {journal} {In practical implementations, measurement
  of the probability is taken at the end of the pulse where decoherence does
  not yet affect the amplitude. In our model, there is no explicit inclusion of
  coherent mechanisms and so, the specific time at which measurement is
  performed is less critical}\ }\BibitemShut {NoStop}%
\bibitem [{\citenamefont {Koppens}\ \emph {et~al.}(2006)\citenamefont
  {Koppens}, \citenamefont {Buizert}, \citenamefont {Tielrooij}, \citenamefont
  {Vink}, \citenamefont {Nowack}, \citenamefont {Meunier}, \citenamefont
  {Kouwenhoven},\ and\ \citenamefont {Vandersypen}}]{koppens06}%
  \BibitemOpen
\bibfield  {journal} {  }\bibfield  {author} {\bibinfo {author} {\bibfnamefont
  {F.~H.~L.}\ \bibnamefont {Koppens}}, \bibinfo {author} {\bibfnamefont
  {C.}~\bibnamefont {Buizert}}, \bibinfo {author} {\bibfnamefont {K.~J.}\
  \bibnamefont {Tielrooij}}, \bibinfo {author} {\bibfnamefont {I.~T.}\
  \bibnamefont {Vink}}, \bibinfo {author} {\bibfnamefont {K.~C.}\ \bibnamefont
  {Nowack}}, \bibinfo {author} {\bibfnamefont {T.}~\bibnamefont {Meunier}},
  \bibinfo {author} {\bibfnamefont {L.~P.}\ \bibnamefont {Kouwenhoven}}, \ and\
  \bibinfo {author} {\bibfnamefont {L.~M.~K.}\ \bibnamefont {Vandersypen}},\
  }\href@noop {} {\bibfield  {journal} {\bibinfo  {journal} {Nature}\ }\textbf
  {\bibinfo {volume} {442}},\ \bibinfo {pages} {766} (\bibinfo {year}
  {2006})}\BibitemShut {NoStop}%
\bibitem [{\citenamefont {Landau}(1932)}]{landau32}%
  \BibitemOpen
  \bibfield  {author} {\bibinfo {author} {\bibfnamefont {L.}~\bibnamefont
  {Landau}},\ }\href@noop {} {\bibfield  {journal} {\bibinfo  {journal}
  {Physics of the Soviet Union}\ }\textbf {\bibinfo {volume} {2}},\ \bibinfo
  {pages} {46} (\bibinfo {year} {1932})}\BibitemShut {NoStop}%
\bibitem [{\citenamefont {Zener}(1932)}]{zener32}%
  \BibitemOpen
  \bibfield  {author} {\bibinfo {author} {\bibfnamefont {C.}~\bibnamefont
  {Zener}},\ }\href@noop {} {\bibfield  {journal} {\bibinfo  {journal} {Proc.
  R. Soc. London A}\ }\textbf {\bibinfo {volume} {137}},\ \bibinfo {pages}
  {696} (\bibinfo {year} {1932})}\BibitemShut {NoStop}%
\bibitem [{\citenamefont {Stueckelberg}(1932)}]{stueckelberg32}%
  \BibitemOpen
  \bibfield  {author} {\bibinfo {author} {\bibfnamefont {E.~C.~G.}\
  \bibnamefont {Stueckelberg}},\ }\href@noop {} {\bibfield  {journal} {\bibinfo
   {journal} {Helvetica Physica Acta}\ }\textbf {\bibinfo {volume} {5}},\
  \bibinfo {pages} {369} (\bibinfo {year} {1932})}\BibitemShut {NoStop}%
\bibitem [{\citenamefont {Foletti}\ \emph {et~al.}(2009)\citenamefont
  {Foletti}, \citenamefont {Bluhm}, \citenamefont {Mahalu}, \citenamefont
  {Umansky},\ and\ \citenamefont {Yacoby}}]{foletti09}%
  \BibitemOpen
  \bibfield  {author} {\bibinfo {author} {\bibfnamefont {S.}~\bibnamefont
  {Foletti}}, \bibinfo {author} {\bibfnamefont {H.}~\bibnamefont {Bluhm}},
  \bibinfo {author} {\bibfnamefont {D.}~\bibnamefont {Mahalu}}, \bibinfo
  {author} {\bibfnamefont {V.}~\bibnamefont {Umansky}}, \ and\ \bibinfo
  {author} {\bibfnamefont {A.}~\bibnamefont {Yacoby}},\ }\href@noop {}
  {\bibfield  {journal} {\bibinfo  {journal} {Nature Phys.}\ }\textbf {\bibinfo
  {volume} {5}},\ \bibinfo {pages} {903} (\bibinfo {year} {2009})}\BibitemShut
  {NoStop}%
\bibitem [{\citenamefont {Hahn}(1950)}]{hahn50}%
  \BibitemOpen
  \bibfield  {author} {\bibinfo {author} {\bibfnamefont {E.~L.}\ \bibnamefont
  {Hahn}},\ }\href@noop {} {\bibfield  {journal} {\bibinfo  {journal} {Phys.
  Rev.}\ }\textbf {\bibinfo {volume} {80}},\ \bibinfo {pages} {580} (\bibinfo
  {year} {1950})}\BibitemShut {NoStop}%
\bibitem [{\citenamefont {Carr}\ and\ \citenamefont {Purcell}(1954)}]{carr54}%
  \BibitemOpen
  \bibfield  {author} {\bibinfo {author} {\bibfnamefont {H.~Y.}\ \bibnamefont
  {Carr}}\ and\ \bibinfo {author} {\bibfnamefont {E.~M.}\ \bibnamefont
  {Purcell}},\ }\href@noop {} {\bibfield  {journal} {\bibinfo  {journal} {Phys.
  Rev.}\ }\textbf {\bibinfo {volume} {94}},\ \bibinfo {pages} {630} (\bibinfo
  {year} {1954})}\BibitemShut {NoStop}%
\bibitem [{\citenamefont {Maestri}\ \emph {et~al.}(2000)\citenamefont
  {Maestri}, \citenamefont {Landau},\ and\ \citenamefont {Paez}}]{maestri00}%
  \BibitemOpen
  \bibfield  {author} {\bibinfo {author} {\bibfnamefont {J.~J.~V.}\
  \bibnamefont {Maestri}}, \bibinfo {author} {\bibfnamefont {R.~H.}\
  \bibnamefont {Landau}}, \ and\ \bibinfo {author} {\bibfnamefont {M.~J.}\
  \bibnamefont {Paez}},\ }\href@noop {} {\bibfield  {journal} {\bibinfo
  {journal} {Am. J. Phys.}\ }\textbf {\bibinfo {volume} {69}},\ \bibinfo
  {pages} {1113} (\bibinfo {year} {2000})}\BibitemShut {NoStop}%
\bibitem [{\citenamefont {Visscher}(1991)}]{visscher91}%
  \BibitemOpen
  \bibfield  {author} {\bibinfo {author} {\bibfnamefont {P.~B.}\ \bibnamefont
  {Visscher}},\ }\href@noop {} {\bibfield  {journal} {\bibinfo  {journal}
  {Comput. Phys.}\ }\textbf {\bibinfo {volume} {5}},\ \bibinfo {pages} {596}
  (\bibinfo {year} {1991})}\BibitemShut {NoStop}%
\bibitem [{\citenamefont {Leforestier}\ \emph {et~al.}(1991)\citenamefont
  {Leforestier}, \citenamefont {Bisseling}, \citenamefont {Cerjan},
  \citenamefont {Feit}, \citenamefont {Friesner}, \citenamefont {Guldberg},
  \citenamefont {Hammerich}, \citenamefont {Jolicard}, \citenamefont
  {Karrlein}, \citenamefont {Meyer}, \citenamefont {Lipkin}, \citenamefont
  {Roncero},\ and\ \citenamefont {Kosloff}}]{leforestier91}%
  \BibitemOpen
  \bibfield  {author} {\bibinfo {author} {\bibfnamefont {C.}~\bibnamefont
  {Leforestier}}, \bibinfo {author} {\bibfnamefont {R.~H.}\ \bibnamefont
  {Bisseling}}, \bibinfo {author} {\bibfnamefont {C.}~\bibnamefont {Cerjan}},
  \bibinfo {author} {\bibfnamefont {M.~D.}\ \bibnamefont {Feit}}, \bibinfo
  {author} {\bibfnamefont {R.}~\bibnamefont {Friesner}}, \bibinfo {author}
  {\bibfnamefont {A.}~\bibnamefont {Guldberg}}, \bibinfo {author}
  {\bibfnamefont {A.}~\bibnamefont {Hammerich}}, \bibinfo {author}
  {\bibfnamefont {G.}~\bibnamefont {Jolicard}}, \bibinfo {author}
  {\bibfnamefont {W.}~\bibnamefont {Karrlein}}, \bibinfo {author}
  {\bibfnamefont {H.-D.}\ \bibnamefont {Meyer}}, \bibinfo {author}
  {\bibfnamefont {N.}~\bibnamefont {Lipkin}}, \bibinfo {author} {\bibfnamefont
  {O.}~\bibnamefont {Roncero}}, \ and\ \bibinfo {author} {\bibfnamefont
  {R.}~\bibnamefont {Kosloff}},\ }\href@noop {} {\bibfield  {journal} {\bibinfo
   {journal} {J. Comp. Phys.}\ }\textbf {\bibinfo {volume} {94}},\ \bibinfo
  {pages} {59} (\bibinfo {year} {1991})}\BibitemShut {NoStop}%
\bibitem [{\citenamefont {Munshi}(2014)}]{opencl14}%
  \BibitemOpen
  \bibinfo {editor} {\bibfnamefont {A.}~\bibnamefont {Munshi}},\ ed.,\
  \href@noop {} {\emph {\bibinfo {title} {The OpenCL Specification, Version:
  2.0}}}\ (\bibinfo  {publisher} {Khronos OpenCL Working Group},\ \bibinfo
  {year} {2014})\BibitemShut {NoStop}%
\end{thebibliography}%

\end{document}